\documentclass{elsart3}

\usepackage{graphicx}
\usepackage{epsfig}
\usepackage{amssymb,amsmath,amscd}

\begin{document}

\begin{frontmatter}

\title{Simulation for KM3NeT using ANTARES-Software}

\author{Sebastian Kuch}

\address{University Erlangen-N\"urnberg, Physikalisches Institut, \\Erwin-Rommel-Str.1, D-91058 Erlangen}

\begin{abstract}
The KM3NeT project is a common European effort for the design of a $\text{km}^3$-scale deep-sea neutrino telescope in the Mediterranean. For the upcoming Design Study simulations have been done using modified ANTARES software. Several concepts and ideas have been tested for their merits and feasibility. 
\end{abstract}

\end{frontmatter}

\section{Introduction}
\label{intro}
The European neutrino telescope experiments \cite{uli}
have joined their efforts in the KM3NeT project \cite{km3net} to design a 
$\text{km}^3$-scale deep-sea neutrino telescope in the Mediterranean. A 
neutrino telescope of these dimensions on the Northern hemisphere, 
complementary to IceCube \cite{icecube}, is necessary for high-energy neutrino 
astronomy.\par
As a first step for the KM3NeT project, a detailed simulation study of 
different detector models and photo-detectors is necessary. 
The software used by the ANTARES collaboration, with some modifications, is 
flexible enough for this task and was used for this study.\par
The modifications include the adaptation to a larger detector and a new 
causality filter for the event reconstruction \cite{zab}. In addition some 
changes were necessary to implement the different photo-detection systems 
presented in this work.\par
Nonetheless the reconstruction and the selection cuts are optimized for 
ANTARES, and are therefore not necessarily optimal for different detector 
designs. However as the results show similar efficiencies for the detector 
models studied in this work and for the ANTARES detector, any reconstruction 
induced inefficiencies must be small.\par
The event sample used contained $10^9$ muon neutrinos with energies distributed
between $10$ and $10^7\,\text{GeV}$ and incident isotropically from the whole 
solid angle. Only charged-current $\nu$N interactions were simulated, and the 
hadronic component of the final state was neglected. A \nuc{40}{K} background rate of 
91 Hz per $\text{cm}^2$ of photocathode area was used, corresponding to 
$40\,\text{kHz}$ for a 10" Photomultiplier.\par
For comparison of different detector models the neutrino effective area was 
calculated as a function of the neutrino energy.
The angular resolution, defined as the median angular deviation
between reconstructed and true neutrino direction, was derived from the 
results, again as a function of energy.\par

\section{The cylindrical storey}
\label{cylOM}
\begin{figure}
\begin{center}
\includegraphics[angle=0,width=5cm]{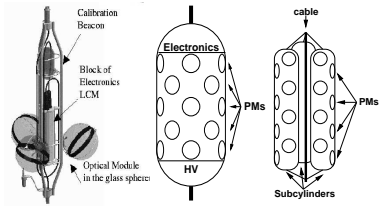}\\
\end{center}
\caption{From left to right: ANTARES storey; Schematic  view of single and triple cylinder layouts, see \protect\cite{nikhef}.}
\label{fig1}
\end{figure}

The different Photomultiplier (PM) configurations (storeys) considered in this work are depicted 
in fig.\ref{fig1}.\par
$\bullet$  ANTARES storey with 3 10" PMs.\par
$\bullet$  Single Cylinder with 35 3" PMs, where each PM is read out individually.\par
$\bullet$  Three smaller cylinders with 12 3" PMs each, hits occuring in one 
of the sub-cylinders, within a predefined time window are added.\par
The overall Photocathode area of these structures is comparable. Small 
PMs generally have a higher quantum efficiency and a better Transit Time Spread
(TTS). 
A disadvantage of the single cylinder structure lies in its vulnerability to 
mechanical stability problems. Additionally individual readout of 35 PMs within
a very confined space might not be feasible. Therefore the triple cylinder 
variant was proposed.  
By adding hits in one sub-cylinder the amount of necessary readout electronics 
is reduced considerably.\par
\begin{figure}[h]
\begin{center}
\includegraphics[angle=270,width=7cm]{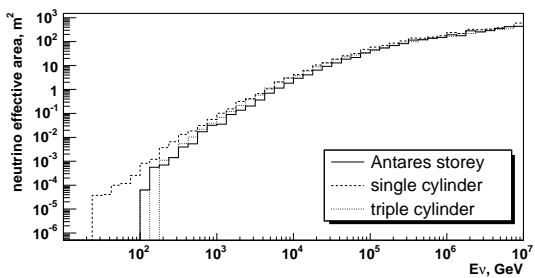}\\
\end{center}
\caption{Neutrino effective areas for a cube grid detector, instrumented with
the different storey structures.}
\label{fig2}
\end{figure}
For comparison these photodetector units where placed in a cubic kilometer grid
with a string and storey distance of $62.5\,\text{m}$ (4913 storeys). Effective
areas calculated for these detectors are shown in fig.\ref{fig2}. Obviously the
single cylinder storey is superior at energies below $1\,\text{TeV}$.\par
The reconstruction requires a certain minimal number of PM hits, a fact that is
especially important at low energies, where the number of signal hits is very 
low. In the single cylinders the high number of individual hits explains the 
good low-energy efficiency.\par
At energies between one and a few hundred TeV, the single and triple cylinder 
storeys are comparable and both superior to the ANTARES storey. For the single 
cylinder this can again be explained by a larger number of hits. The reason in 
case of the triple cylinder is the requirement of local coincidences in one 
storey and/or hits with high amplitudes. Coincident hits in one of the triple 
cylinders are added and therefore transformed into a high amplitude hit. 
The angular resolution is very similar for all three cases, with a common 
overall median of $0.07^{\circ}$.\par 

\section{Inhomogeneous geometries}
\label{geom}
\begin{figure}[h]
\begin{center}
\includegraphics[angle=270,width=3cm]{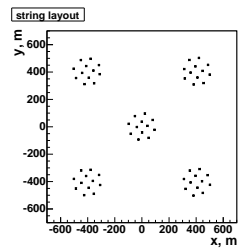}\\
\end{center}
\caption{Cluster layout: 5 clusters with 12 strings each, 4980 storeys.}
\label{fig3}
\end{figure}

The efficiency of a detector at low energies is correlated to the distance 
between the PMs. A possible way to combine low- and high-energy performance is 
to use clusters of densely instrumented strings as shown in fig.\ref{fig3}. 
Low-energy muons have a high detection probability inside the 
clusters, while high-energy muons have a chance to hit several of the clusters,
thus seeing a large instrumented volume.\par
\begin{figure}[h]
\begin{center}
\includegraphics[angle=270,width=7cm]{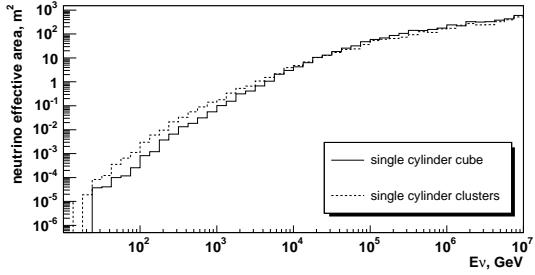}\\
\end{center}
\caption{Neutrino effective areas for a cluster detector instrumented with
single cylinder storeys.}
\label{fig4}
\end{figure}

\begin{figure}[h]
\begin{center}
\includegraphics[angle=270,width=7cm]{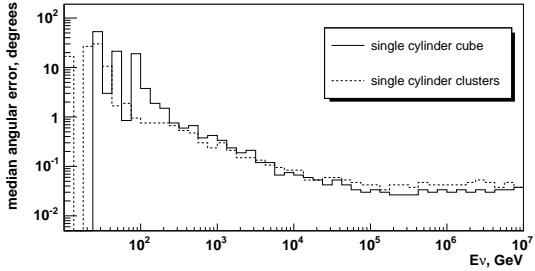}\\
\end{center}
\caption{Angular resolution for a cluster detector instrumented with the single
cylinder storeys.}
\label{fig5}
\end{figure}

As expected, effective area (fig.\ref{fig4}), and angular resolution 
(fig.\ref{fig5}) are significantly better than for the cube at energies below 
$20\,\text{TeV}$. At higher energies the performance is about 20\% worse.\par
 

Above approximately $1\,\text{TeV}$ the muon range in water exceeds the 
dimensions of the instrumented volume of a $\text{km}^3$-scale detector.
Starting from this energy, most of the muons will enter the detector from the
outside. Therefore the cross section area of the detector starts to become more
important than a densely instrumented volume.\par
\begin{figure}
\begin{center}
\includegraphics[angle=0,width=6cm]{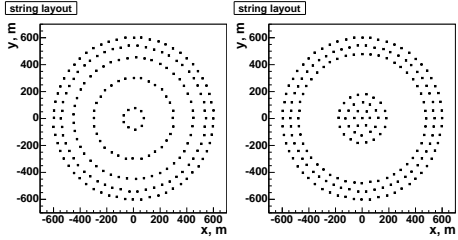}\\
\end{center}
\caption{Seafloor layout of ring1 (left) and ring2 (right) detector: 200(199) 
strings, 5000(4975) storeys.}
\label{fig6}
\end{figure}

In order to exploit this effect the design of a ring-shaped detector with a 
densely instrumented boundary was studied. Examples of string layouts for such 
detectors are shown in fig.\ref{fig6}.\par
\begin{figure}
\begin{center}
\includegraphics[angle=270,width=7cm]{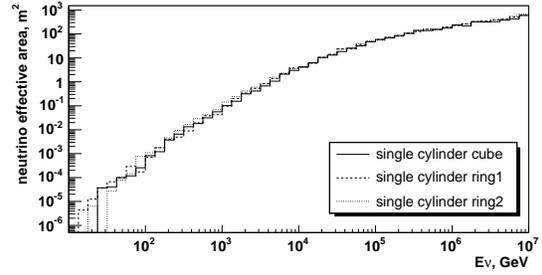}\\
\end{center}
\caption{Neutrino effective areas for ring detectors instrumented with single 
cylinder storeys.}
\label{fig7}
\end{figure}

Effective areas for the rings, as shown in fig.\ref{fig7}, are slightly higher 
at low energies, due to the denser storey spacing in the ring. The 
angular resolution is very similar to the homogeneous case.

\section{Conclusions}
\label{Conclusions}
Several promising concepts of PM configurations and geometries for the design 
of the Mediterranean $\text{km}^3$ neutrino detector were considered.\par
Through simulations it was shown that the use of many small PMs in pressure 
cylinders can provide better performance as conventional large hemispherical 
PMs. Inhomogeneous geometries have been shown to increase efficiency at low 
energies with only small losses at high energies, while drastically reducing 
the number of necessary strings.\par
The simulations also demonstrated that ring geometries are a possible way to 
reduce the number of strings without any loss of performance.\par
The decision for a definite configuration has to depend on the physics 
priorities of the project, as the performance of the different detector 
concepts depends on the neutrino energy. Further studies with dedicated 
software for a $\text{km}^3$-scale detector are absolutely necessary to further
clarify the results presented here, since the software used for this work is 
optimized for the ANTARES detector.\par

\end{document}